\documentclass[11pt,a4paper]{article}
\usepackage{graphicx,axodraw}
\usepackage{amsmath,amsxtra,amssymb,latexsym, amscd,amsfonts}
\usepackage[mathscr]{eucal}

\makeindex
\begin{document}
\parindent=1.05cm 
\setlength{\baselineskip}{12truept} \setcounter{page}{1}
\makeatletter
\renewcommand{\@evenhead}{\@oddhead}
\renewcommand{\@oddfoot}{}
\renewcommand{\@evenfoot}{\@oddfoot}\renewcommand{\thesection}{\arabic{section}.}
\renewcommand{\theequation}{\thesection\arabic{equation}}
\@addtoreset{equation}{section}
\title{}
\date{} 
\maketitle
\vspace{-4cm}
\begin{center}
{\large \bf ELIMINATING ON THE DIVERGENCES \\OF THE PHOTON SELF -
ENERGY DIAGRAM\\ IN (2+1) DIMENSIONAL QUANTUM ELECTRODYNAMICS}
\vspace{12pt}\\ 
\textbf{Nguyen Suan Han} \vspace{6pt}\\
\it{Department of physics, College of Science, Hanoi National
University - Viet nam}
\vspace{9pt}\\
 \textbf{Nguyen Nhu Xuan}\footnote{Email:
fieldtheory2876@gmail.com }
 \vspace{6pt}\\
\it {Department of physics, Le Qui Don Technical University - Viet
nam}
\end{center}
\begin{center}
\vspace{0.3cm} \small \noindent {\bf Abstract}
\end{center}
\begin{quote}
\it The divergence of the photon self - energy diagram in spinor
quantum electrodynamics in $(2+1)$ dimensional space time- $(QED_3)$
is studied by the Pauli - Villars regularization and dimensional
regularization. Results obtained by two different methods are
coincided if the gauge invariant of theory is considered carefully
step by step in these calculations.
\end{quote}
\vspace{0.3cm}
\leftskip0cm 
\section{INTRODUCTION}
\hspace{0.7cm} It is well known that the gauge theories in $(2+1)$
dimensional space time though  super-renormalizable theory $[1]$,
showing up inconsistences already at one loop, arising from the
regularization procedures adopted to evaluate ultraviolet divergent
amplitudes such as the photon self-energy in $QED_3$. In the latter,
if we use dimensional regularization $[2]$ the photon is induced a
topological mass in contrast with the result obtained through the
Pauli-Villars scheme $[3]$, where the photon remains massless when
we let the auxiliary mass go to infinity. Other side this problem is
important for constructing quantum field theory with low
dimensional modern.\vspace{6pt}\\
\indent This report is devoted to show up the inconsistencies not
arising in $QED_3$, if the gauge invariance of theory is considered
carefully step by step in those calculations by above methods of
regularization for the photon self - energy diagram. The paper is
organized as follows. In the second section the photon self-energy
is calculated by the dimensional regularization. In the third
section this problem is done by the Pauli-Villars method. Finally,
we draw our conclusions.
\section{DIMENSIONAL REGULARIZATION}
\hspace{0.7cm} In this section, we calculate the photon self -
energy diagram in $ QED_3 $ given by $(\textbf{Fig 1})$
\begin{center}
\begin{picture} (200,30)(-20,0)
\Photon(0,0)(40,0){5}{2.5} \Vertex(40,0){2}\Vertex(80,0){2}
\ArrowArc(60,0)(20,0,180) \ArrowArc(60,0)(20,180,360)
\Photon(80,0)(120,0){5}{2.5} \LongArrow(12,-10)(25,-10)
\LongArrow(92,-10)(105,-10) \Text(20,12)[]{k} \Text(100,12)[]{k}
\Text(60,28)[]{p} \Text(60,-28)[]{p-k}
\end{picture}\vspace{1.2cm}\\
{\textbf{Fig.1}:\emph{The photon self - energy diagram}}.
\vspace{0.2cm}
\end{center}
\hspace{0.7cm} Following the standard notation, this graph is
corresponding to the formula:
\begin{equation}\label{2.1}
\Pi_{\mu\nu}(k)=\frac{ie^2}{(2\pi)^{\frac{3}{2}}}\int d^3p
Tr\left[\gamma_\mu\frac{\hat{p}+m}{p^2-m^2+i\epsilon}
\gamma_\nu\frac{\hat{p}-k+m}{(p-k)^2-m^2+i\epsilon}\right].
\end{equation}
\indent In dimensional regularization scheme, we have to make the
change :
\begin{equation}\label{2.2}
\int \frac{d^3p}{(2\pi)^{\frac{3}{2}}}\rightarrow\mu^\epsilon \int
\frac{d^np}{(2\pi)^{\frac{n}{2}}},
\end{equation}
where $\epsilon=3-n, \mu $ is some arbitrary mass scale which
is introduced to preserve dimensional of system.\vspace{6pt}\\
\indent Make to shift $p$ by $p+\frac{1}{2}k $, the expression
$(2.1)$ has the form :
\begin{equation}\label{2.3}
    \begin{split}
\Pi_{\mu\nu}(k)&=ie^2\mu^\epsilon\int\frac{d^np}{(2\pi)^{\frac{n}{2}}}
\left[\gamma_\mu\frac{\left(\hat{p}+\frac{1}{2}\hat{k}\right)+m}
{\left(p+\frac{1}{2}k\right)^2-m^2+i\epsilon}
\gamma_\nu\frac{\left(\hat{p}-\frac{1}{2}\hat{k}\right)+m}
{(p-\frac{1}{2}k)^2-m^2+i\epsilon}\right]\\
 &= ie^2\mu^\epsilon\int\frac{d^np}{(2\pi)^{\frac{n}{2}}}
\int_0^1 dx \frac{P(m)}{\left[m^2-p^2+(x^2-x)k^2\right]^2},
    \end{split}
\end{equation}
with
\begin{equation}\label{2.4}
\begin{split}
P(m)=&2\Bigr\{m^2g_{\mu\nu}+2p_\mu p_\nu +(1-2x)p_\nu k_\mu
+2(x^2-x)k_\mu k_\nu\\
&-g_{\mu\nu}\left[p^2+(1-2x)pk+(x^2-x)k^2\right]
-im\epsilon_{\mu\nu\alpha}k^\alpha\Bigr\}.
\end{split}
\end{equation}
In the expression $(2.3)$, we have used Feynman integration
parameter$[5]$.\vspace{6pt}\\
\indent Neglecting the integrals that contain the odd terms of p in
$P(m)$ which will vanish under the symmetric integration in p. Then
we have
\begin{equation}\label{2.5}
\begin{split}
\Pi_{\mu\nu}(k)=&2ie^2\mu^\epsilon\int_0^1 dx\int
\frac{d^np}{(2\pi)^{\frac{n}{2}}}\times \\
&\left\{\frac{2p_\mu p_\nu}{(p^2-a^2)^2}-\frac{2x(1-x)k_\mu
k_\nu}{(p^2-a^2)^2}+\frac{2x(1-x)k^2g_{\mu\nu}}{(p^2-a^2)^2}-\frac{g_{\mu\nu}}{p^2-a^2}
-\frac{im\epsilon_{\mu\nu\alpha}k^\alpha}{(p^2-a^2)^2}\right\}\\
=&2ie^2\mu^\epsilon\int_0^1 dx\int\frac{d^np}{(2\pi)^{\frac{n}{2}}}
\Biggr\{\frac{2p_\mu p_\nu}{(p^2-a^2)^2}-\frac{g_{\mu\nu}}{(p^2-a^2)}\\
&+\left[\frac{2x(1-x)(k^2g_{\mu\nu}-k_\mu
k_\nu)}{(p^2-a^2)^2}\right]
-\frac{im\epsilon_{\mu\nu\alpha}k^\alpha}{(p^2-a^2)^2}\Biggr\}.
\end{split}
\end{equation}
\indent To carry out separating $\Pi_{\mu\nu}(k)$ into three terms
$\Pi_{\mu\nu}(k)=\Pi_{1\mu\nu}(k)+\Pi_{2\mu\nu}(k)+\Pi_{3\mu\nu}(k)$
and using the following formulae of the dimensional regularization :
\begin{align}
I_o=&\int\frac{d^np}{(2\pi)^{\frac{n}{2}}}\frac{1}{(p^2-a^2)^\alpha}
=
\frac{i(-\pi)^{\frac{n}{2}}{}}{(2\pi)^{\frac{n}{2}}}\frac{\Gamma\left(\alpha-\frac{n}{2}\right)}{\Gamma(\alpha)}
\frac{1}{(-\alpha^2)^{\left(\alpha-\frac{n}{2}\right)}},\label{2.6}\\
I_{\mu\nu}=&\int\frac{d^np}{(2\pi)^{\frac{n}{2}}}\frac{p_\mu
p_\nu}{(p^2-a^2)^\alpha} =
\frac{g_{\mu\nu}(-\alpha^2)}{\left(\alpha-1-\frac{n}{2}\right)}I_o
,\label{2.7}\\
\Gamma\left(2-\frac{n}{2}\right)=&\left(1-\frac{n}{2}\right)\Gamma\left(1-\frac{n}{2}\right),
\hspace{12pt}\Gamma(2)=\Gamma(1)=1,\label{2.8}
\end{align}
 we obtain:
 \begin{align}
\Pi_{1\mu\nu}(k)&=2ie^2\mu^\epsilon\int_0^1dx\int\frac{d^np}{(2\pi)^{\frac{n}{2}}}
\left[\frac{2p_\mu p_\nu}{\left(p^2-a^2\right)^2}
-\frac{g_{\mu\nu}}{\left(p^2-a^2\right)}\right]=0,\label{2.9}\\
\Pi_{2\mu\nu}(k)&=2ie^2\mu^\epsilon\int_0^1 2x(1-x)
\left(k^2g_{\mu\nu}-k_\mu k_\nu\right)dx
\int\frac{d^np}{(2\pi)^{\frac{n}{2}}}\frac{1}{\left(p^2-a^2\right)^2} \notag\\
&=\frac{e^2\mu^\epsilon\left(k_\mu k_\nu-k^2g_{\mu\nu}\right)}
{(2\pi)^{-1/2}}\int_0^1dx\frac{x(1-x)}{\left[m^2-x(1-x)k^2\right]^{1/2}},\label{2.10}\\
\Pi_{3\mu\nu}(k)&=2e^2m\epsilon_{\mu\nu\alpha}\mu^\epsilon k^\alpha
\int_0^1 dx\int\frac{d^np}{(2\pi)^{\frac{n}{2}}}\times
\frac{1}{\left(p^2-a^2\right)^2}\notag\\
&=e^2m\epsilon_{\mu\nu\alpha}\mu^\epsilon k^\alpha
\frac{i\sqrt{\pi}}{\sqrt{2}}\int_0^1 dx
\frac{1}{\left[m^2-x(1-x)k^2\right]^{1/2}}.\label{2.11}
 \end{align}
\indent From the expression $(2.10)$, we are easy to see that
$\Pi_{2\mu\nu}(0)=0$. So the final result, we find :
 \begin{equation}\label{2.12}
\Pi_{\mu\nu}(k)_{k^2=0}=\left[\Pi_{1\mu\nu}(k)+\Pi_{2\mu\nu}(k)+
\Pi_{3\mu\nu}(k)\right]|_{k^2=0} \Rightarrow
\Pi_{\mu\nu}(k)_{k^2=0}=\Pi_{3\mu\nu}(k)_{k^2=0}\neq 0.
\end{equation}
\indent The expression $(2.12)$ talk to us that in the dimensional
regularization method the photon have additional mass that is
differential from zero, even its momentum  equal zero. \vskip6pt
\indent In the next section, we will study this problem by
Pauli-Villars regularization method.
\section{PAULI-VILLARS REGULARIZATION}
\hspace{0.7cm} Pauli-Villars regularization consists in replacing
the singular Green's functions of the massive free field with the
linear combination $[4]$ :
\begin{equation}\label{3.1}
\Delta(x)\rightarrow reg_M\Delta(m)=\Delta(m)+\sum_i c_i\Delta(M_i).
\end{equation}
Here the symbol  $\Delta_c(m)$  stands for the Green's function of
the field of mass m, and the symbol $\Delta(M_i)$ are auxiliary
quantities representing Green's function of fictitious fields with
mass $M_i$, while  $c_i$ are certain coefficients satisfying special
conditions. These conditions are chosen so that the regularized
function $reg\Delta(x;m)$ considered in the configuration
representation turns out to be sufficiently regular in the vicinity
of the light cone, or (what is equivalent) such that the function
$\bar{\Delta}(p;m)$ in the momentum
representation falls off sufficiently fast in the region of large  $|p|^2$.\vspace{6pt}\\
\indent On the base of Pauli - Villars regularization we calculate
the polarization tensor operator in $QED_3$. For the vacuum
polarization tensor we find the following expression :
\begin{equation}\label{3.2}
 \Pi_{\mu\nu}^M(k)= \frac{ie^2}{(2\pi)^{3/2}}\sum_i^{n_f}c_i
 \int d^3p \frac{Tr\left[\gamma_\mu\left(M_i+\hat{p}-\frac{1}{2}\hat{k}\right)
 \gamma_\nu\left(M_i+\hat{p}-\frac{1}{2}\hat{k}\right)\right]}
 {\left[M_i^2+\left(p-\frac{1}{2}k\right)^2\right]\times
 \left[M_i^2+\left(p+\frac{1}{2}k\right)^2\right]},
\end{equation}
with
$c_o=1;M_o=m;M_i=m\lambda_i$;$\sum_{i=0}^{n_f}c_i=0;\sum_{i=0}^{n_f}c_iM_i=0$
$(i=1,2,...n_f)$.\vspace{6pt}\\
\indent For simplicity, but without loss of generality,  we may
choose both the electron mass and two mass of auxiliary fields to be
positive, the coefficients $\lambda_i$ ultimately go to infinity to
recover the original theory.
\begin{equation}\label{3.3}
\Pi_{\mu\nu}^M(k)=\frac{ie^2}{(2\pi)^{3/2}}\sum_i^{n_f}c_i \int d^3p
\int_0^1dx\frac{P(M_i)}{\left[M_i^2-p^2+(x^2-x)k^2\right]^2} .
\end{equation}
\begin{equation}\label{3.4}
\begin{split}
P(M_i)=&2\Bigr\{M_i^2g_{\mu\nu}+2p_\mu p_\nu +(1-2x)p_\nu k_\mu
+2(x^2-x)k_\mu k_\nu \\
&-g_{\mu\nu}[p^2+(1-2x)pk+(x^2-x)k^2]-iM_i
\epsilon_{\mu\nu\alpha}k^\alpha\Bigr\} .
\end{split}
\end{equation}
Neglecting the integrals that contain the odd terms of p in $P(M_i)$
we get:
\begin{equation}\label{3.5}
\begin{split}
\Pi_{\mu\nu}^ M (k)&=\frac{ie^2}{(2\pi)^{3/2}}\sum_i^{n_f}c_i \int
d^3p \int_0^1dx\times \\
&\frac{2\left\{M_i^2g_{\mu\nu}+2p_\mu p_\nu +2(x^2-x)k_\mu k_\nu
-g_{\mu\nu}p^2-2g_{\mu\nu}\left(x^2-x\right)k^2-iM_i
\epsilon_{\mu\nu\alpha}k^\alpha\right\}}{\left[M_i^2-p^2+(x^2-x)k^2\right]^2}.
\end{split}
\end{equation}
 The expression  $\Pi_{\mu\nu}^M(k)$ can be written in
the form:
\begin{equation}\label{3.6}
\Pi_{\mu\nu}^M(k)=\left(g_{\mu\nu}-\frac{k_\mu
k_\nu}{k^2}\right)\Pi_1^M(k^2)+im\epsilon_{\mu\nu\alpha}k^\alpha
\Pi_2^M(k^2)+g_{\mu\nu}\Pi_3^M(k^2).
\end{equation}
\indent Set $a_i^2=M_i^2+(x^2-x)k^2=M_i^2-x(1-x)k^2$, we have:
\begin{align}
\Pi_1^M(k^2)=&4ie^2\sum_{i=0}^{n_f}c_i\int_0^1x(1-x)dx\int
\frac{d^3p}{(2\pi)^{3/2}}\times\frac{1}{\left(a_i^2-p^2\right)^2},\label{3.7}\\
\Pi_2^M(k^2)=&-\frac{2ie^2}{m}\sum_{i=0}^{n_f}c_iM_i\int_0^1dx
\int\frac{d^3p}{(2\pi)^{3/2}}\times\frac{1}{\left(a_i^2-p^2\right)^2},\label{3.8}
\end{align}
\begin{equation}\label{3.9}
\Pi_3^M(k^2)= 2ie^2\sum_{i=0}^{n_f}c_i\left[\int_0^1dx
\int\frac{d^3p}{(2\pi)^{3/2}}\frac{1}{\left(a_i^2-p^2\right)^2}+2\int_0^1dx
\int\frac{d^3p}{(2\pi)^{3/2}}\frac{P^2}{\left(a_i^2-p^2\right)^2}
\right].
\end{equation}
If we carry out these integrations in the momentum space, it is
straightforward
to arrive: $\Pi_3^M(k^2)=0$ as expected by the gauge invariance.\vspace{6pt}\\
\indent Here comes the crucial point: we can't blindly take only one
auxiliary field with $ M= \lambda m$ as usual; this choice is
missionary conditions
$\sum_{i=0}^{n_f}c_i=0;\sum_{i=0}^{n_f}c_iM_i=0$ must be matched.
This is possible only fixing $ \lambda=1 $ . Thus, the number of
regulators must be at leat two, otherwise we can't get the
coefficients $\lambda_i $ becoming arbitrarily large. So, let us
take : $c_1=\alpha-1;c_2=-\alpha;c_j=0$ when $j>2$.\vspace{6pt}\\
\indent Where the parameter $\alpha$ can assume any real value
except zero and the unity, so that. condition $(3.3)$ is satisfied.
For
 $\lambda_1, \lambda_2\rightarrow \infty $ and apply it to $(3.6)$. To pay attention
 $\Gamma(1/2)=\sqrt{\pi}$, we have
\begin{equation}\label{3.10}
\begin{split}
\Pi_1^M=&4ie^2\sum_{i=0}^{n_f}c_i\int_0^1 dx x(1-x)
\int\frac{d^3p}{(2\pi)^{3/2}}\times
\frac{1}{\left(a_i^2-p^2\right)^2} \\
=&-\frac{e^2k^2}{(2\pi)^{-1/2}}\int_0^1 dx x(1-x)
\left[\frac{c_o}{(a_o^2)^{1/2}}+\frac{c_1}{(a_1^2)^{1/2}}
+\frac{c_2}{(a_2^2)^{1/2}}\right],
\end{split}
\end{equation}
where
\begin{equation}\label{3.11}
a_o^2=m^2-x(1-x)k^2;a_1^2=\lambda_1m_1^2-x(1-x)k^2;
a_2^2=\lambda_2m_2^2-x(1-x)k^2.
\end{equation}
Thus , when $\lambda_1, \lambda_2\rightarrow \infty$ :
\begin{equation}\label{3.12}
    \Pi_1^M(k^2)\rightarrow \Pi_1(k^2)=-\frac{e^2k^2}{(2\pi)^{-1/2}}
    \int_0^1dx x(1-x)\frac{1}{\left[m^2-x(1-x)k^2\right]^{1/2}},
\end{equation}
and consequently, $ \Pi_1(0)=0 $.\vspace{6pt}\\
\indent From $(3.8)$, we have :
\begin{equation}\label{3.13}
\begin{split}
\Pi_2^M(k^2)=&\frac{e^2}{4m\pi}\int_0^1dx
\Biggr\{\frac{m}{\left[m^2-x(1-x)k^2\right]^{1/2}}\\
&+\frac{(\alpha-1)M_1}{\left[M_1^2-x(1-x)k^2\right]^{1/2}}
-\frac{(\alpha-1)M_2}{\left[M_2^2-x(1-x)k^2\right]^{1/2}}\Biggr\}.
\end{split}
\end{equation}
\indent Taking the limit $\lambda_1, \lambda_2\rightarrow \pm\infty$
(depending on couplings $ c_1 $ and $ c_2 $ having the same sign or
different sign $\lambda\rightarrow +\infty$ or $\lambda\rightarrow
-\infty$), for photon momentum  k=0, yields:
\begin{itemize}
\item if $\lambda_1\rightarrow +\infty; \lambda_2\rightarrow
\infty $: the couplings $ c_1, c_2 $ have the different sign, and $
\alpha<0$ or $ \alpha >0 $, to
\begin{equation}
\Pi_2(0)=0.
\end{equation}
\item if $\lambda_1\rightarrow +\infty; \lambda_2\rightarrow
-\infty$: the couplings $ c_1, c_2 $ have the same sign, and
  $ 0<\alpha<1 $
\begin{equation}
\Pi_2(0)=\frac{e^2}{\sqrt{2}m(\pi)^{3/2}}(1+\alpha-1+\alpha)
=\frac{2e^2\alpha}{\sqrt{2}m(\pi)^{3/2}}.
\end{equation}
From the results $(3.14)$ and $(3.15)$, we can be written them in
the form
\begin{equation}
\Pi_2(0)=\frac{\alpha e^2}{\sqrt{2}m \pi^{3/2}}(1-s),
\end{equation}
with $s=sign \left(1-\frac{1}{\alpha} \right)$.\\
\end{itemize}
\hspace{0.7cm} It is obvious that, from $(3.16)$, we saw: if $
0<\alpha<1 $ and $ s=-1 $ the couplings $ c_1, c_2 $ have the same
sign $ \Pi_2(0)\neq 0 $; in this case photon requires a topological
mass, proportional to $\Pi_2(0) $, coming from proper insertions of
the antisymmetry sector of the vacuum polarization tensor in the
free photon propagator. If we assume that $ \alpha $ is outside this
range
$(0,1)$ and $ c_1 $ and $c_2 $ have opposite signs and  $ \Pi_2(0)=0$.\vspace{6pt}\\
\indent We then conclude that this arbitrariness $\alpha $ reflects
in different values for the photon mass. The new parameter $s$ may
be identified with the winding number of homologically nontrivial
gauge transformations and also appears in lattice regularization $[7]$.\vspace{6pt}\\
\indent Now we face another problem: which value of $ \alpha $ leads
to the correct photon mass? A glance at equation $(3.9)$ and we
realize that $ \Pi_2(k^2) $ is ultraviolet finite by naive power
counting. We were taught that a closed fermion loop must be
regularized  as a whole so to preserve gauge invariance. However
having done that we have affected a finite antisymmetric piece of
the vacuum polarization tensor and, consequently, the photon mass.
The same reasoning applies when, using Pauli-Villars regularization,
we calculate the anomalous magnetic moment of the electron; again,
if care is not taken, we may arrive at a wrong physical result.\vspace{6pt}\\
\indent  In order to get of this trouble we should pick out the
value of $\alpha$ that cancels the contribution coming from the
regulator fields. From expression $(3.16)$, we easily find that this
occurs for $ (c_1=c_2)$ because in this case the signs of the
auxiliary masses are opposite, in account of condition $(3.3)$. From
$(3.16)$, we obtain $\Pi_2(0)=\frac{e^2}{\sqrt{2}m \pi^{3/2}}$, in
agreement with the other approach already mentioned. We should
remember that Pauli Villars regularization violated party symmetry
$(2+1)$ dimensions. Nevertheless, for this particular choice
$\alpha$, this symmetry is restored as regulator mass get larger and
larger.
\begin{center}
\begin{tabular}{|l|l|l|l|}
  \hline
 & $\alpha< 0, \alpha > 1$ & $0 < \alpha < 1$ & photon mass
    \\ \hline
  $c_1$ and $c_2$ opposite sign & $\Pi_2(0)= 0$ &   & equal zero \\ \hline
  $c_1$ and $c_2$ same sign  &  & $\Pi_2(0)\neq$ 0 & unequal zero \\ \hline
  $c_1= c_2;\alpha=\frac{1}{2}$ &  &  & $\Pi_2(0)= -\frac{e^2}{\sqrt{2}m \pi^{3/2}}$ \\
  \hline
\end{tabular}
\end{center}
\section{CONCLUSION}
  \hspace{1cm}In depending on sign of the couplings $c_1$
and $c_2$, same and opposite sign the Pauli-Villars regularization
give a result $\Pi_2(0)=\frac{\alpha e^2}{\sqrt{2}m
\pi^{3/2}}(1-s)$, where $s=sign \left(1-\frac{1}{\alpha}\right)$.
Results obtained by regularization Pauli-Villars and dimensional
methods are coincided if the gauge invariance of theory is
considered carefully step by step in these calculations. When
$c_1=c_2;\alpha=\frac{1}{2}$, the expressions obtained by the Pauli
- Villars and the dimensional method have same results
$\Pi_2(0)=\frac{e^2}{\sqrt{2}m \pi^{3/2}}$ in $QED_3$ in agreement
with the other approaches for these problems $[6]$. This work was
supported by Vietnam National Research Programme in National
Sciences N406406.

\end{document}